\documentclass[%
reprint,
amsmath,amssymb,
aps,
prl,
floatfix
]{revtex4-2}

\usepackage{graphicx}
\usepackage{siunitx}  
\usepackage{booktabs}
\usepackage{hyperref}

\begin{document}

\title{Colloidal Suspensions can have Non-Zero Angles of Repose below the Minimal Value for Athermal Frictionless Particles}

\author{Jesús Fernández}
\author{Loïc Vanel}%
\author{Antoine Bérut}%
 \email{Corresponding author: antoine.berut@univ-lyon1.fr}
\affiliation{Université Lyon 1, CNRS, ILM, UMR 5306, Villeurbanne, France}%

\date{\today}

\begin{abstract}

We investigate the angle of repose $\theta_r$ of dense suspensions of colloidal silica particles ($d = \SI{2}{\micro\meter}$ to \SI{7}{\micro\meter}) in water-filled microfluidic rotating drum experiments, to probe the crossover between the thermal (colloidal) and athermal (granular) regimes. For the smallest particles, thermal agitation promotes slow creep flows, and piles always flatten completely regardless of their initial inclination angle, resulting in $\theta_r = 0$. Above a critical particle size, piles of colloids stop flowing at a finite angle of repose, which increases with particle size but remains below the minimal value expected for athermal frictionless granular materials: $0~<~\theta_r~<~\theta_\mathrm{ath}~\approx~\SI{5.8}{\degree}$. We quantify the arrest dynamics as a function of the gravitational Péclet number $\mathrm{Pe}_g$, which characterizes the competition between particle weight and thermal agitation. Our measurements are consistent with a recent rheological model~\cite{Billon2023}, in which the arrested state stems from a crossover between glass and jamming transitions as the granular pressure in the pile increases relative to the thermal pressure.

\end{abstract}


\maketitle

Unlike simple liquids, many complex fluids, such as colloidal glasses and gels, concentrated emulsions, pastes, and granular materials, exhibit a yield stress below which they behave as solids and do not flow~\cite{larson1999, hunter2012, coussot2014, bonn2017}. In granular media, a classical macroscopic manifestation of this yield stress is the fact that the free surface of a pile can maintain a stable slope under gravity~\cite{andreotti2013granular}. This arrested state is characterized by the angle of repose $\theta_r$, whose value comes from the sum of two main contributions: the geometric interlocking of the particle network, which sets a minimal repose angle $\theta_\mathrm{ath} \approx \SI{5.8}{\degree}$~\cite{Peyneau2008}, and interparticle friction, which typically increases the value of $\theta_r$ to $\sim \SI{30}{\degree}$~\cite{Zhou2002, Pohlman2006}. This description successfully captures the behavior of athermal granular systems, and although most materials have a repose angle $\theta_r \geq \SI{15}{\degree}$~\cite{ReviewAngleofRepose}, it has been demonstrated that $\theta_\mathrm{ath} \approx \SI{5.8}{\degree}$ can be reached experimentally with a suspension of frictionless silica particles of diameter $d \approx \SI{20}{\micro\meter}$~\cite{clavaud2017revealing,Perrin2019}. However, the situation is completely different when particles become sufficiently small to be influenced by Brownian motion, yet still heavy enough to sediment and form a well-defined pile. For dense suspensions of colloidal particles with diameter $d \leq \SI{2}{\micro\meter}$, thermal agitation from the surrounding fluid is sufficient to promote slow creep flows that always flatten the pile over time~\cite{berut2019brownian, lagoin2024effects}. Therefore, these systems cannot sustain any stress, and have a zero angle of repose $\theta_r = \SI{0}{\degree}$.

The transition between the colloidal ($\theta_r = \SI{0}{\degree}$) and granular ($\theta_r \geq \theta_\mathrm{ath}$) limits remains experimentally unexplored and thus an open question~\cite{ikeda2012unified}. A recent phenomenological model has been proposed by Billon \textit{et al.}~\cite{Billon2023} to describe the flow–arrest dynamics in this intermediary regime for thermally agitated suspensions of hard-sphere particles. This model uses the framework of pressure-imposed rheology~\cite{Boyer2011,Guazzelli2018}, in which the granular pressure $\Pi$ is a control parameter and the volume fraction $\phi$ adjusts itself in response to an imposed $\Pi$. It considers contributions of both the thermal (glass) and athermal (jamming) transitions to the yield stress~\cite{Ikeda2013}. The model predicts a discontinuous evolution of the quasi-static friction coefficient of the suspension $\mu_{J=0}$~\footnote{Here, $J=0$ denotes the limit of vanishing viscous number $J$, which can be approached when the shear-rate of the suspensions is close to zero. The viscous number is defined as $J=\eta \dot{\gamma}/\Pi$ (where $\eta$ is the fluid viscosity, $\dot{\gamma}$ the shear rate, and $\Pi$ the granular pressure).} as the dimensionless pressure ${\tilde{\Pi}} = d^3\Pi/k_\mathrm{B}T$ is varied (where $d$ is the particle diameter and $k_\mathrm{B} T$ the thermal energy). This discontinuity is associated with the crossing of the glass transition because the volume fraction of the system $\phi$ increases with $\tilde{\Pi}$. At low pressure, $\phi < \phi_G$ (where $\phi_G \approx 0.58$ denotes the critical volume fraction at which the glass transition occurs), therefore the suspension has no yield stress~\cite{bonn2017}. Accordingly, both $\mu_{J=0}$ and the angle of repose $\theta_r = \tan^{-1}\left(\mu_{J=0}\right)$ are equal to zero in this case. Then, at a critical pressure ${\tilde{\Pi}^*}$, at which $\phi = \phi_G$, a yield stress suddenly appears and $\mu_{J=0}$ becomes non-zero. Beyond this threshold, the friction coefficient and the corresponding angle of repose $\theta_r$ increase with $\tilde{\Pi}$ toward the jamming limit (defined by $\phi = \phi_J \approx 0.64$~\cite{Guazzelli2018}), revealing intermediate non-zero values in the thermal to athermal crossover. Although this scenario has already been observed numerically~\cite{Wang2015}, no experimental validation has been reported so far.

In this letter, we measure the inclination angle at which sedimented piles of colloidal particles stop flowing after thermal creep, in order to investigate how their final angle of repose $\theta_r$ evolves with the ratio between particle weight and thermal agitation. Experimentally, we tune this ratio by varying the particle diameter from \SI{2}{\micro\meter} to \SI{7}{\micro\meter}, which modifies their gravitational Péclet number:
\begin{equation}
\mathrm{Pe}_g = \frac{m g d}{k_\mathrm{B} T},
\end{equation}
where $m = \pi d^3 \Delta\rho / 6$ is the buoyant particle mass, $g$ the gravitational acceleration, $d$ the particle diameter, and $k_\mathrm{B} T$ the thermal energy. We report that for small particles (low $\mathrm{Pe}_g$), piles relax completely to a flat free surface, exhibiting no angle of repose. Then, beyond a critical particle size threshold, piles stop flowing at a non-zero angle, which increases with $\mathrm{Pe}_g$ but remains smaller than the minimal athermal value $\theta_\mathrm{ath}$. Finally, we compare our experimental results with the predictions of the model~\cite{Billon2023}, and find a good agreement with a single fitting parameter.

\begin{table*}[!t]
\centering
\caption{Experimental particle diameters $d$ (with standard deviation reported by the manufacturer) and corresponding gravitational Péclet numbers $\mathrm{Pe}_g$.}
\label{tab:Pe_values}
\begin{tabular}{cccccccc}
\toprule
$d$ [\si{\micro\meter}] \,
& \, 1.93\,$\pm$\,0.05 \,
& \, 2.12\,$\pm$\,0.06 \,
& \, 2.40\,$\pm$\,0.04 \,
& \, 2.83\,$\pm$\,0.06 \,
& \, 2.96\,$\pm$\,0.07 \,
& \, 3.97\,$\pm$\,0.12 \,
& \, 7.00\,$\pm$\,0.15 \, \\
\midrule
$\mathrm{Pe}_g$
& 15\,$\pm$\,2
& 21\,$\pm$\,2
& 35\,$\pm$\,2
& 68\,$\pm$\,6
& 81\,$\pm$\,8
& 264\,$\pm$\,32
& 2548\,$\pm$\,218 \\
\bottomrule
\end{tabular}
\end{table*}

\begin{figure}[ht]
    \centering
    \includegraphics[width=\columnwidth]{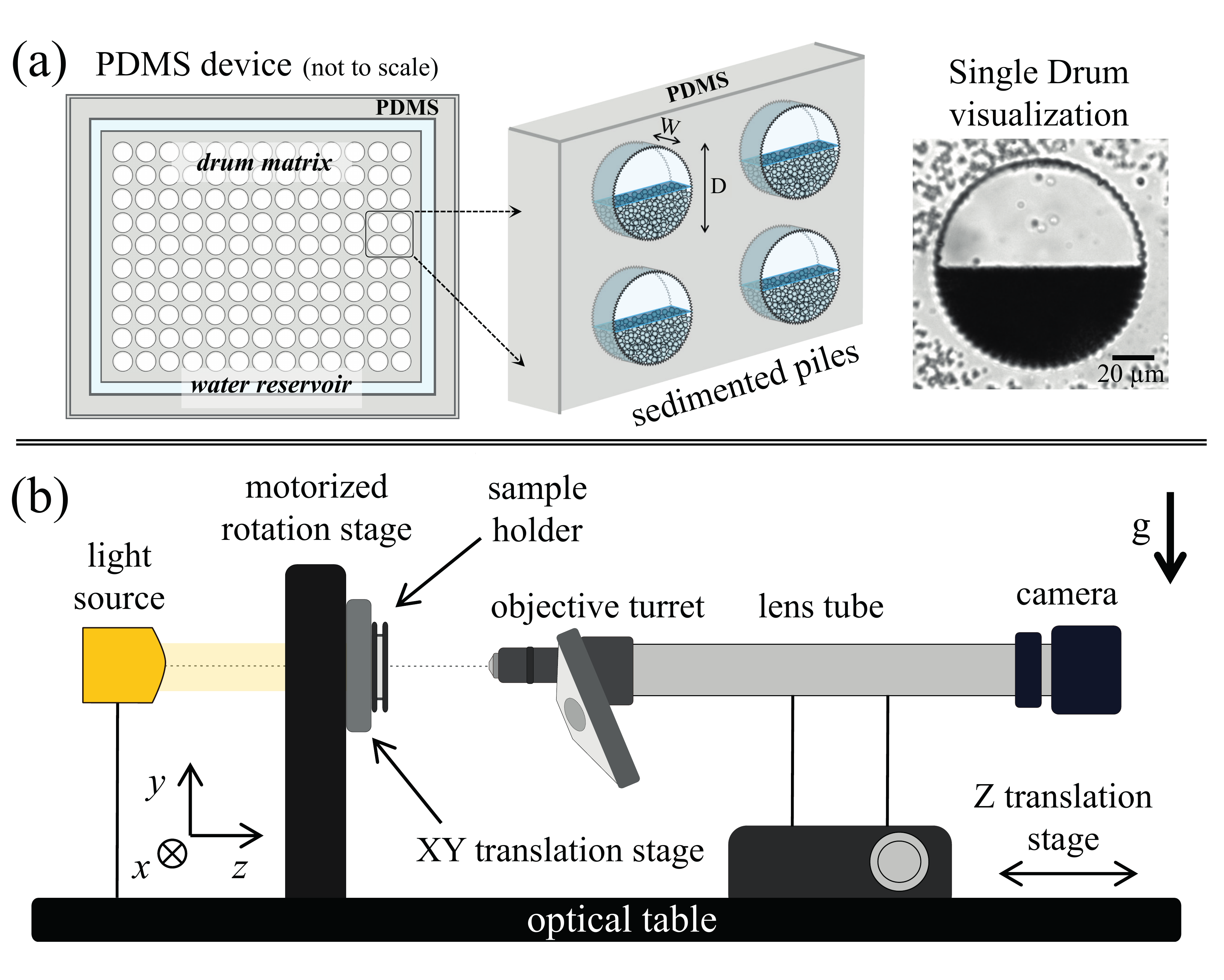} 
    \caption{Experimental set-up. (a) Microfluidic drum sample. Left: schematic drawing of a PDMS drum array with the surrounding water moat (not to scale). Center: 3D renderings of four drums half-filled with the colloidal suspension sedimented at their bottom. Right: microscope picture of a single drum. (b) Schematic drawing of the sample mounted on a vertical rotation stage and coupled to a horizontal microscope for simultaneous imaging of twenty drums (the arrow $g$ represents the direction of gravity).}
    \label{fig:FigI}
\end{figure}

Our experiments are carried out in a microfluidic rotating-drum platform designed for long (up to one month) and stable observation of dense particle piles. Figure~\ref{fig:FigI} illustrates the experimental setup: micrometer-sized PDMS drums (diameter $D = 100~\si{\micro\meter}$, width $W = 50~\si{\micro\meter}$) are filled with suspensions of monodisperse silica particles (from \href{https://microparticles.de/}{\textit{microParticles GmbH}}) in ultrapure water (\textit{Millipore Direct-Q 3 UV}) and sealed with a glass slide, following the protocol described in~\cite{Fernandez2025}. The design incorporates a surrounding millimetric water moat that minimizes sample evaporation and triangular grooves patterned along the drum walls (pattern width $\sim$5~\si{\micro\meter}), which provide a rough boundary that prevents particle sliding. The device is mounted vertically on a motorized rotation stage (\textit{Newport URB100CC}) coupled to a horizontal microscope with a 10$\times$ objective (\textit{Olympus MPLFLN 10$\times$/0.30}) and a CCD camera (\textit{Basler acA2440-75um}), all placed on an optical table. This set-up allows simultaneous imaging of 20 drums with a spatial resolution of 0.30~\si{\micro\meter}/pixel at frequency up to \SI{75}{\hertz}. 

When placed vertically, the particles sediment under gravity and form a dense pile at the bottom of each drum (typically occupying $\sim\qty{50}{\percent}$ of the volume). Silica particles acquire a negative surface charge in water through the dissociation equilibrium $\equiv$SiOH $\rightleftharpoons$ $\equiv$SiO$^-$ + H$^+$, generating strong short-range electrostatic repulsion~\cite{israelachvili2011intermolecular}, which results in a Debye length $\lambda_D\approx \qty{50}{\nano\meter}$ in ultrapure water~\cite{lagoin2024effects}. Under the low confining pressure imposed by the microfluidic geometry, this repulsion prevents direct solid–solid contacts. Before each measurement, the sample is rotated at \SI{90}{\degree\per\second} for \SI{1}{\hour} to redisperse the particles, after which the system is left undisturbed for \SI{30}{\minute} until a reproducible initial sedimented state is reached.

The angle of repose $\theta_r$ is measured through a flow-to-arrest protocol: the piles are first put into flow by tilting the drums rapidly (within less than \qty{1}{\second}) to a chosen initial inclination $\theta_{\mathrm{start}}$, and the subsequent evolution of the free surface is monitored until the dynamics cease. Over $2000$ images are acquired with a logarithmic frame rate. The instantaneous pile angle $\theta(t)$ is extracted using automated contour detection and averaged over the twenty drums (see \hyperref[image_analysis]{Sup. Mat.}). We define the angle of repose $\theta_r$ as the smallest angle for which $\theta(t)$ shows no measurable evolution over a time comparable to the full relaxation time (\textit{i.e.}, the time needed for the pile to flow from $\theta_\mathrm{start}$ to $\theta_r$). This definition corresponds to the lower angle of repose commonly measured for granular materials in rotating drum experiments~\cite{Liu2005}. We apply this protocol to piles of seven particle sizes, $d \approx \qtyrange{2}{7}{\micro\meter}$, spanning gravitational Péclet numbers $\mathrm{Pe}_g \approx \qtyrange{15}{2500}{}$ (Table~\ref{tab:Pe_values}). Each measurement is repeated twice on the same device and once on an independent device to confirm the absence of sample degradation and ensure reproducibility.

\begin{figure}[ht]
    \centering
    \includegraphics[width=\columnwidth]{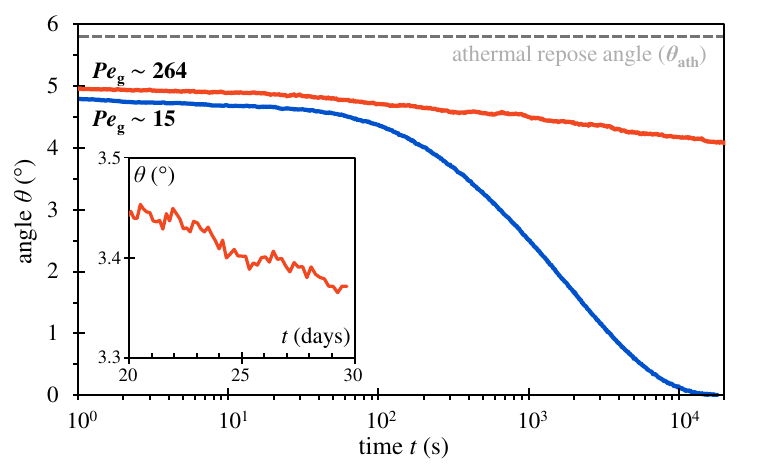} 
    \caption{Free-surface creep dynamics of gravitationally sedimented piles tilted to \SI{5}{\degree} for $\mathrm{Pe}_g \approx 15$ and $\mathrm{Pe}_g \approx 264$ (inset: long-time evolution for $\mathrm{Pe}_g \approx 264$).}
    \label{fig:FigII}
\end{figure}

Figure~\ref{fig:FigII} shows the free-surface flow dynamics of piles with particle diameter $d = \qty{1.93(0.05)}{\micro\meter}$ ($\mathrm{Pe}_g \approx 15$) and $d = \qty{3.97(0.12)}{\micro\meter}$ ($\mathrm{Pe}_g \approx 264$) following a tilt to $\theta_\mathrm{start}=\SI{5}{\degree}$. At this inclination, the piles lie below the minimal angle that geometric interlocking can sustain under gravity $\theta_\mathrm{ath} \approx \SI{5.8}{\degree}$, and therefore gravity-driven flows are precluded~\footnote{When the flow is initiated above $\theta_\mathrm{ath}$, it proceeds in two stages: an initial fast gravity driven flow, followed by a slow thermally activated creep flow~\cite{berut2019brownian,lagoin2024effects}.}. Instead, the observed dynamics is dominated by a slow thermally activated creep, which strongly depend on the gravitational Péclet number~\cite{berut2019brownian}. Thus, the time required to reach their arrested state differs by several orders of magnitude between the two particle sizes. For $\mathrm{Pe}_g \approx 15$, the pile fully relaxes within $\sim \qty{5}{\hour}$, whereas for $\mathrm{Pe}_g \approx 264$, creep persists at extremely slow rate for over one month without reaching an angle of repose (see inset in Fig.~\ref{fig:FigII}). These observations highlight the intrinsic difficulty of directly verifying the arrested state of piles at high $\mathrm{Pe}_g$ within experimentally accessible times.

\begin{figure}[ht]
    \centering
    \includegraphics[width=\columnwidth]{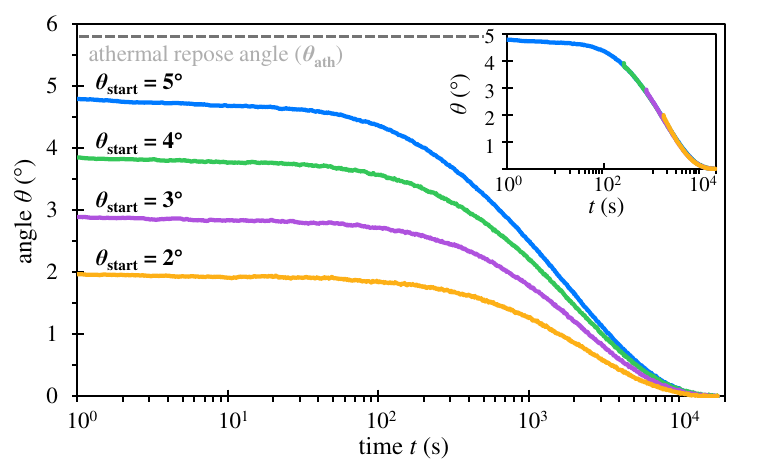} 
    \caption{Free-surface creep dynamics of sedimented piles initially inclined at different $\theta_{\mathrm{start}}$ for $\mathrm{Pe}_g\approx 15$. Inset: same data collapsed onto a single trajectory. The trajectory with $\theta_\mathrm{start}=\qty{5}{\degree}$ is taken as the reference curve, and each trajectory $\theta(t)$ with $\theta_\mathrm{start}<\qty{5}{\degree}$ is shifted in time by an offset $t_\mathrm{shift}$, chosen to achieve the best collapse onto the reference curve.}
    \label{fig:FigIII}
\end{figure}

To circumvent experimentally inaccessible long relaxation times and measure the angle of repose at high Péclet numbers, we trigger the flow at progressively smaller initial angles ($\theta_{\mathrm{start}} < \SI{5}{\degree}$), until reaching a value of $\theta_\mathrm{start}$ for which the pile ceases to flow within the experimental time frame. By doing so, we assume that the final angle of repose $\theta_r$ is independent of the initial inclination angle $\theta_\mathrm{start}$. This hypothesis seems reasonable since the creep flow is slowly driven by thermal agitation, and the pile microstructure is not expected to retain a memory of the deformation history. Although this cannot be confirmed experimentally for high $\mathrm{Pe}_g$ due to extremely long creep duration~\footnote{Note that an extrapolation of the creep dynamics for $\mathrm{Pe}_g \approx 264$, assuming no early arrest at a finite angle, indicates that it would require several tens of years for the pile to reach \SI{0}{\degree}, which we cannot measure experimentally.}, we have verified it for $\mathrm{Pe}_g\approx 15$, where the complete relaxation dynamics can be measured. Figure~\ref{fig:FigIII} shows that for different values of $\theta_{\mathrm{start}}$ the final arrested state is always the same ($\theta_r=0$). Moreover, the relaxation curves $\theta(t)$ all collapse onto a single master trajectory after a simple time shift (see inset of Fig.~\ref{fig:FigIII}), demonstrating that the creep rate at a given time $\dot{\theta}(t)$ is governed only by the value of the pile angle at the same time $\theta(t)$ and does not depend on the flow history. Finally, any potential aging effects along the trajectory appear negligible on the overall timescale of the creep~\cite{Fernandez2025}.

\begin{figure}[ht]
    \centering
    \includegraphics[width=\columnwidth]{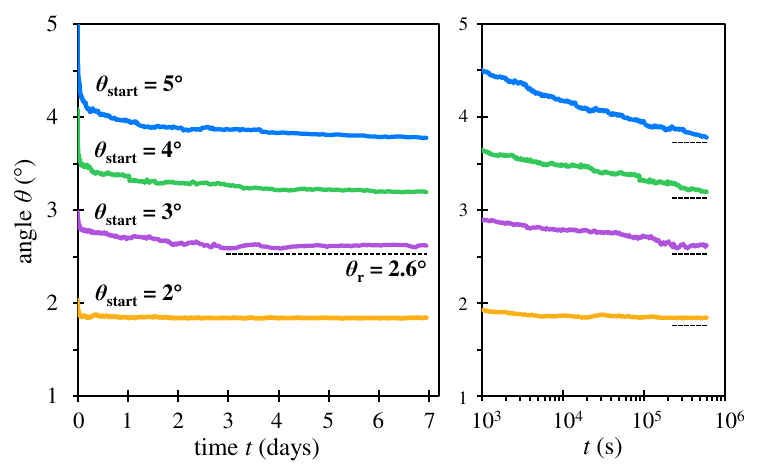} 
    \caption{Free-surface creep dynamics of sedimented piles initially inclined at different angle $\theta_{\mathrm{start}}$ for $\mathrm{Pe}_g\approx 264$, shown in linear scale (left) and semi-log scale (right). Dashed lines indicate the horizontal. The piles with $\theta_{\mathrm{start}} = \SI{3}{\degree}$ stop flowing after $\sim 3$ days, arresting at $\theta_r \approx \SI{2.6}{\degree}$.}
    \label{fig:FigIV}
\end{figure}

Figure~\ref{fig:FigIV} shows the temporal evolution of the piles with $\mathrm{Pe}_g \approx 264$ following tilts to $\theta_{\mathrm{start}}$ ranging from \SI{5}{\degree} to \SI{2}{\degree}, each monitored over one week. Flows initiated at \SI{5}{\degree} and \SI{4}{\degree} do not reach an arrested state during this time period. Piles tilted to $\theta_{\mathrm{start}} = \SI{3}{\degree}$ stop flowing after approximately three days, reaching a finite angle of repose $\theta_r \approx \SI{2.6}{\degree}$. Consistently, when the piles are tilted to \SI{2}{\degree}, no measurable relaxation is observed over a week. These observations provide direct experimental evidence for a non-zero angle of repose, intermediate between the vanishing angle observed at low $\mathrm{Pe}_g$ (as in Fig.~\ref{fig:FigIII}) and the minimal expected value $\theta_\mathrm{ath} \approx \SI{5.8}{\degree}$ in the limit of granular systems without thermal creep.

\begin{figure}[ht]
    \centering
    \includegraphics[width=\columnwidth]{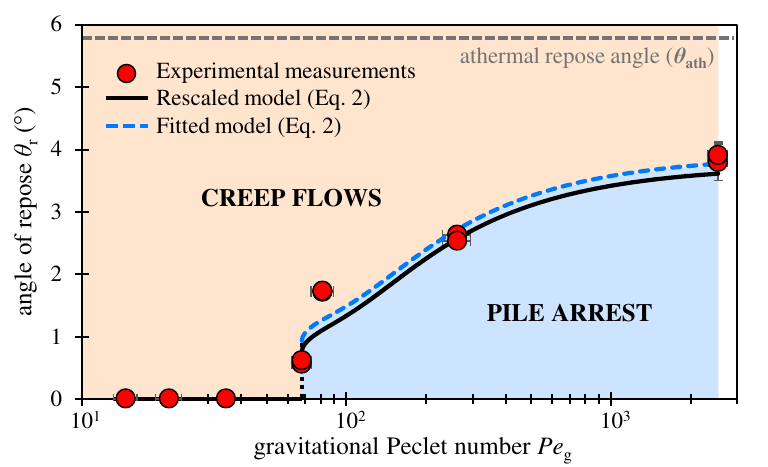} 
    \caption{Measured angles of repose $\theta_r$ as a function of the gravitational Péclet number $\mathrm{Pe}_g$. Both solid and dashed lines show predictions from the rheological model (Eq.~\ref{eq:model}): the black solid curve shows the effect of linearly rescaling the dimensionless pressure to match the gravitational Péclet number with a single fitting parameter ($\mathrm{Pe}_g= \alpha \tilde{\Pi}$), while the blue dashed curve also includes $Y_G$ and $Y_G^\prime$ as free fitting parameters.}
    \label{fig:FigV}
\end{figure}

We have systematically applied this angle-sweeping protocol to piles prepared with each particle size studied (table~\ref{tab:Pe_values}). This allows us to quantify how the angle of repose $\theta_r$ evolves with the gravitational Péclet number $\mathrm{Pe}_g$. Results are shown in Figure~\ref{fig:FigV}. For $\mathrm{Pe}_g \leq 35$, we measure $\theta_r=0$, indicating the absence of yield stress. Then, as $\mathrm{Pe}_g$ increases above a critical value $\mathrm{Pe}_g^*\approx 68$, the angle of repose $\theta_r$ becomes non-zero. It increases from $\SI{0.6}{\degree}$ at $\mathrm{Pe}_g \approx 68$ to $\SI{3.7}{\degree}$ at $\mathrm{Pe}_g \approx 2548$, corresponding to yield stresses in the range $10^{-4}$ to $10^{-3}$ \si{\pascal}~\footnote{The value of the yield stress can be estimated from the shear stress on a particle at the top of the pile at the repose angle: $\sigma_\mathrm{yield} \approx \frac{4mg}{\pi d^2}\sin (\theta_r)$}. Additionally, we have verified that $\theta_r$ is not affected by wall confinement or the drum geometry (see \hyperref[boundary_confinement]{Sup. Mat.}).

Finally, we compare our measurements with the predictions of the rheological model introduced by Billon \textit{et al.}~\cite{Billon2023}. In this model, the angle of repose $\theta_r$ is analytically expressed as a function of the dimensionless pressure $\tilde{\Pi}$:
\begin{equation}
\tan \left( \theta_r \right) = \mu_{J=0} = \frac{\sigma}{\Pi} = \frac{Y_G}{\tilde{\Pi}} + \frac{Y_G^\prime}{\tilde{\Pi}}
\frac{\left[\phi - \phi_G\right]^{\beta_G}}
{\left(\phi_J - \phi\right)},
\label{eq:model}
\end{equation}
where $\phi$ is the volume fraction, $\phi_G=0.575$ and $\phi_J = 0.64$ denote respectively the glass and jamming volume fractions for hard spheres, and $Y_G = 0.38$, $Y_G^\prime = 0.17$, and $\beta_G = 0.6$ are dimensionless parameters derived from the expression of the thermal contribution to the shear stress $\sigma$~\cite{Ikeda2013}.
Then, to fully determine $\theta_r$, one needs to express the volume fraction $\phi$ as a function of the dimensionless pressure $\tilde{\Pi}$. Following Billon~\cite{BillonPhD}, this can be achieved by inverting a modified Carnahan-Starling equation of state for hard spheres~\footnote{Although our particles are not strictly hard spheres, the small size of their repulsive layer ($\lambda_D \approx \qty{50}{\nano\meter}$) compared to their diameter ($d = \qtyrange{2}{7}{\micro\meter}$) justifies the hard sphere approximation.}:
\begin{equation}
\tilde{\Pi} \left( \phi \right) = \frac{6}{\pi} \phi \frac{\phi_J}{\phi_J - \phi} \frac{1+\phi+\phi^2 - 7.5\phi^3}{(1-\phi)^2} .
\end{equation}

Although the dimensionless pressure $\tilde{\Pi}$ used in the model and the experimental Péclet number $\mathrm{Pe}_g$ are two distinct physical quantities, they both quantify the competition between confinement stress and thermal agitation. In the model, $\tilde{\Pi}$ quantifies the ratio between the granular pressure $\Pi$ (pressure applied to the grains) and the thermal stress $k_\mathrm{B} T/d^3$. In our experiments, the granular pressure arises from the particles weight. Thus, we expect $\Pi \propto \frac{mg}{d^2}$, which yields directly $\tilde{\Pi} \propto \mathrm{Pe}_g$. Note that a vertical gradient of pressure is expected in the pile under gravitational loading. However, previous experiments have shown that the creep flow is localized in the topmost particles layers~\cite{berut2019brownian}. Therefore, we consider that the relevant pressure governing the dynamics is the local pressure in the top flowing layers, which is constant in average. We fit the model (Eq~\ref{eq:model}) to our data with a single free parameter $\alpha$, defined by $\mathrm{Pe}_g= \alpha \tilde{\Pi}$. The result of the model obtained with the best fitting parameter ($\alpha = 2.4$) is plotted in Figure~\ref{fig:FigV} (black solid line). This minimal rescaling shows a good agreement between the model's prediction and our experimental measurements.

A slightly better agreement can be obtained by allowing $Y_G$ and $Y_G^\prime$, whose value are empirical~\cite{Ikeda2013}, to become free fitting parameters. The result of this fit is shown in Figure~\ref{fig:FigV} (blue dashed line). In this case, we obtain the best agreement with $Y_G = 0.47 \pm 0.10$ and $Y_G^\prime = 0.18 \pm 0.01$, whose values remain very close to those used in the model: $Y_G = 0.38$, $Y_G^\prime = 0.17$~\cite{Billon2023}.

Within the framework of the model, the emergence of a non-zero angle of repose at a critical dimensionless pressure $\tilde{\Pi}^*$ corresponds to the crossing of the glass transition for the system when $\phi = \phi_G$. To verify the consistency of this hypothesis, we have measured by confocal imaging the volume fraction of a sedimented pile of fluorescent \qty{2.03}{\micro\meter} particles ($\mathrm{Pe}_g \approx 18$) in our microfluidic drums. This suspension, which exhibits no yield stress ($\theta_r = 0$), has a volume fraction $\phi = \qty{0.558(0.007)}{}$, below the expected value for the glass transition $\phi_G \approx 0.58$. The proximity of this volume fraction $\phi$ to $\phi_G$ suggests that the system may cross the glass transition when the pressure is increased (i.e. for larger particles). Although we do not directly probe a glass transition at $\mathrm{Pe}_g^*\approx 68$ and cannot determine experimentally whether the transition is truly discontinuous or not, our results appear consistent with the underlying assumptions of the model.

In the context of granular materials, thermal agitation can be seen as analogous to mechanical vibrations, which are known to induce logarithmic relaxations of the angle of repose toward \SI{0}{\degree} in dry piles~\cite{Jaeger1989}, and to suppress the yield stress in macroscopic suspensions~\cite{Hanotin2012}. However, this analogy remains qualitative, as the equivalence between thermal agitation and mechanical vibrations is still an open question. Some models~\cite{Mehta1992,Luck2004} consider a statistical effect of the vibrations, equivalent to an effective temperature, but predict that a weakly dilatant pile should always have a zero repose angle when vibrated, in contradiction with our observations. Other models consider that the main effect of the mechanical vibration is to reduce the frictional behavior of the pile~\cite{Linz1994,Garat2022}, which is irrelevant here since our particles are frictionless. Recent experiments in inclined planes~\cite{Gaudel2016}, later confirmed by numerical simulations~\cite{Gaudel2019}, have shown that a granular pile can have a non-zero repose angle that decreases with the amplitude of an applied horizontal vibration, but no angle below $\theta_\mathrm{ath}$ has been reported. To our knowledge, no macroscopic model predicts how the angle of repose of a granular material evolves under mechanical vibrations, and our measurements could provide a useful comparison point for the development of such models.

\textit{Conclusion.} In this letter, we have experimentally demonstrated that dense colloidal suspensions in a pressure-imposed configuration exhibit slow creep flows below the minimal angle of repose expected for granular suspensions ($\theta_\mathrm{ath} \approx \SI{5.8}{\degree}$), and can arrest at a finite angle $0 < \theta_r < \theta_\mathrm{ath}$. These flows are driven by thermal agitation of the surrounding fluid and depend on the weight of the particles. By systematically tilting sedimented piles to progressively smaller angles, we were able to probe their arrested state despite extremely long relaxation times. We have mapped the angle of repose $\theta_r$ as a function of the ratio between particle weight and thermal agitation, quantified by the gravitational Péclet number $\mathrm{Pe}_g$. The observed dependency of $\theta_r$ with $\mathrm{Pe}_g$ is in good agreement with the predictions of a phenomenological model in the pressure-imposed rheology framework, which considers contributions of both the thermal (glass) and athermal (jamming) transitions to describe the arrest dynamics~\cite{Billon2023}. Our results thus provide experimental evidence for non-zero angles of repose in the thermal-to-athermal crossover of dense suspensions, thereby bridging the rheology of colloidal and granular materials.

\begin{acknowledgments}
The authors acknowledge the support of the French Agence Nationale de la Recherche (ANR) under grant ANR-21-CE30-0005 (MicroGraM). All data supporting the findings of this study are openly available~\cite{dataFernandez}.
\end{acknowledgments}

\vspace{-5mm}

\bibliographystyle{apsrev4-2}
\bibliography{biblio} 

\newpage

\appendix

\onecolumngrid

\section{Supplemental Material}

In this Supplemental Material, we describe the procedure used to measure the pile angle from image sequences, and we present additional experiments on the relaxation dynamics of the pile to demonstrate that the final angle of repose $\theta_r$ is independent of the drum geometry.

\subsection{Image analysis to measure the pile angle}
\label{image_analysis}

\begin{figure}[ht!]
    \centering
    \includegraphics[width=\linewidth]{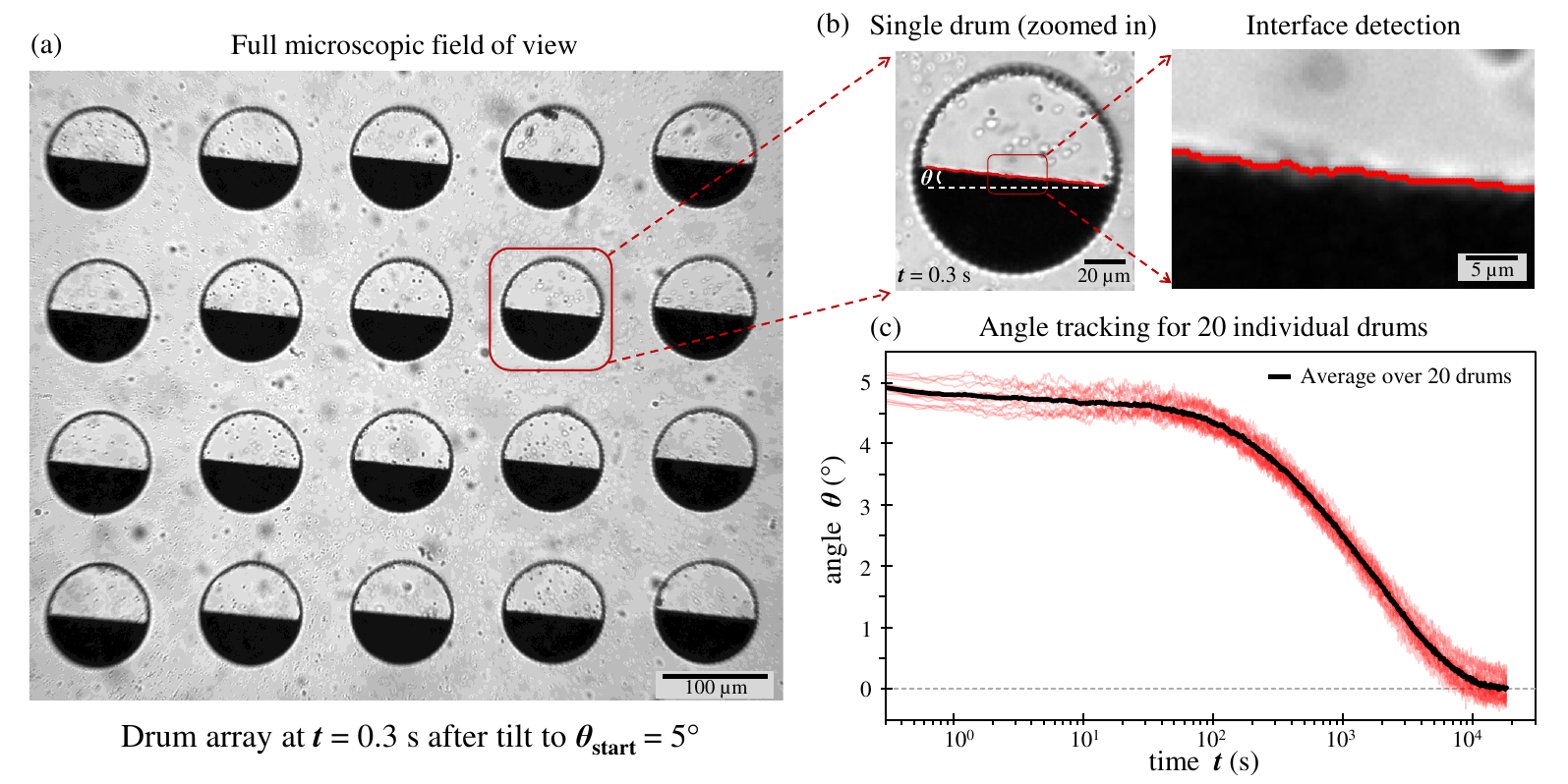}
    \caption{
    Measurement of the instantaneous pile angle $\theta(t)$ for particles of diameter $d = \qty{1.93(0.05)}{\micro\meter}$ ($\mathrm{Pe}_g \approx 15$), in drums of diameter $D = \qty{100}{\micro\meter}$ and width $ W = \qty{50}{\micro\meter}$.
    (a) Typical image showing the array of 20 drums within the field of view. 
    (b) Interface detection: the contour of the sedimented pile is identified (red line, left), and a zoomed view (right) shows the interface resolved at the pixel level.
    (c) Time evolution of the angle $\theta(t)$ for the 20 individual drums (red lines) and their average (black line).}
    \label{fig:S1}
\end{figure}

Figure~\ref{fig:S1} illustrates the procedure used to measure the instantaneous angle $\theta(t)$ during the relaxation process of the sedimented piles. The analysis is performed on a sequence of images acquired after the tilt (occuring at $t = 0$). A typical frame showing the array of 20 drums within the field of view is displayed in Fig.~\ref{fig:S1} (a).

For each image, the drums are identified and processed individually. First, the drums are detected by image correlation. Then, for each drum, contour of the sedimented pile is detected using an automated edge-detection algorithm based on contrast differences between the dense particle phase and the surrounding fluid (namely, ``contour finding'' from the Python library Scikit-image~\cite{skimage}). The upper boundary of this contour defines the free surface of the pile at the pixel level, see Fig.~\ref{fig:S1} (b). This interface is then fitted by a straight line, excluding the vicinity of the side walls to avoid boundary effects. The slope of this fitted line directly provides the instantaneous angle $\theta_i(t)$ for each drum.

This procedure is repeated for all drums and all times. The reported angle $\theta(t)$ is obtained by averaging over the 20 drums for each frame. The individual time traces and their average are shown in Fig.~\ref{fig:S1}(c), demonstrating that the averaged value provides a robust measure of $\theta(t)$.

The spatial resolution of the images is \qty{0.323}{\micro\meter} per pixel, and the detected interface typically spans about 300 pixels. The dispersion of $\theta$ across the 20 drums is characterized by a standard deviation of approximately \qty{0.4}{\degree}, while the uncertainty on the mean angle (standard error of the mean) is about \qty{0.1}{\degree}.

\subsection{Effects of boundary conditions and confinement}
\label{boundary_confinement}

In rotating drum experiments with macroscopic frictional granular materials, geometric confinement is known to influence both the flow dynamics and the shape of the free surface of the pile. In particular, the drum width $W$ and diameter $D$ are well-established geometric parameters: increasing lateral confinement (small $W/d$) typically enhances the angle of repose, while finite curvature effects (small $D/d$) lead to deviations from a flat free surface~\cite{gdr2004dense}.

Here, we examine whether similar effects arise in piles made of microscopic particles by systematically varying the drum dimensions $W$ and $D$ relative to the particle size $d$.

\begin{figure}[ht!]
    \centering
    \includegraphics[width=\columnwidth]{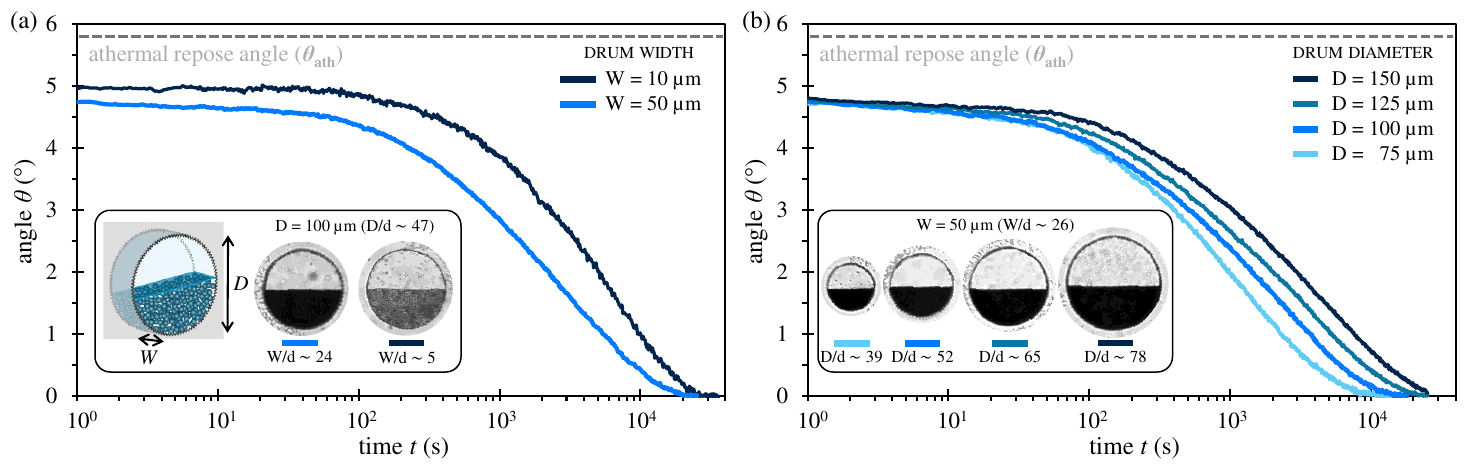} 
    \caption{Free-surface flow dynamics under different confinement conditions.
(a) Effect of lateral confinement $W$: time evolution of the pile angle $\theta(t)$ following a tilt to $\theta_{\mathrm{start}} = \qty{5}{\degree}$ for particles of diameter $d = \qty{2.12(0.06)}{\micro\meter}$ ($\mathrm{Pe}_g \approx 21$), in drums of fixed diameter $D = \qty{100}{\micro\meter}$ and widths $W = \qty{10}{\micro\meter}$ and $\qty{50}{\micro\meter}$.
Inset: schematic of the drum geometry and images of the sedimented piles for each geometry (before the initial tilting).
(b) Effect of drum diameter $D$: time evolution of the pile angle $\theta(t)$ following a tilt to $\theta_{\mathrm{start}} = \qty{5}{\degree}$ for particles of diameter $d = \qty{1.93(0.05)}{\micro\meter}$ ($\mathrm{Pe}_g \approx 15$), in drums of fixed width $W = \qty{50}{\micro\meter}$ and diameters $D = \qtyrange{75}{150}{\micro\meter}$.
Inset: images of the sedimented piles for each geometry (before the initial tilting).}
    \label{fig:confinement}
\end{figure}

Figure~\ref{fig:confinement} shows the free-surface flow dynamics under different confinement conditions. Panel (a) presents variations in the drum width $W$ at fixed $D = \qty{100}{\micro\meter}$, for particles of diameter $d = \qty{2.12(0.06)}{\micro\meter}$ ($\mathrm{Pe}_g \approx 21$). Panel (b) shows analogous measurements for variations in the drum diameter $D$ at fixed $W = \qty{50}{\micro\meter}$, with $d = \qty{1.93(0.05)}{\micro\meter}$ ($\mathrm{Pe}_g \approx 15$). In all cases, the flow is initiated by a tilt to $\theta_{\mathrm{start}} = \qty{5}{\degree}$.

For these small Péclet numbers ($\mathrm{Pe}_g \approx 15$ and $\mathrm{Pe}_g \approx 21$), the full relaxation of the pile can be resolved within accessible experimental times. We find that, across all explored ratios of $W/d$ and $D/d$, the piles can relax through creep down to $\theta_r = 0$. Confinement does not affect the final arrested state of the pile, but instead influences the relaxation dynamics toward it. In particular, decreasing $W$ delays the relaxation without modifying the final angle of repose $\theta_r$ (Fig.~\ref{fig:confinement} (a)). Similarly, larger $D$ leads to longer relaxation times, consistent with the larger volume of material involved in the flow, without modifying $\theta_r$ (Fig.~\ref{fig:confinement} (b)). In all cases, the free surface of the pile remains flat during the relaxation dynamics (see full data available in~\cite{dataFernandez}). Importantly, the experimental conditions explored in the main text of the article (fixed $W = \qty{50}{\micro\meter}$ and $D = \qtyrange{100}{125}{\micro\meter}$, with $d = \qtyrange{2}{7}{\micro\meter}$) fall within the same ranges of $W/d$ and $D/d$ investigated here, supporting the conclusion that confinement does not influence the final arrested state in those experiments.

This behavior contrasts with that of frictional granular systems, where confinement can modify the angle of repose through wall-induced friction~\cite{gdr2004dense}. In our case, interparticle interactions are dominated by short-range repulsion. Indeed, Atomic Force Microscopy (AFM) measurements indicate that silica particles interact repulsively with both PDMS and glass surfaces used for the drums walls, under the low-load conditions relevant to our experiments. These observations, consistent with previous reports~\cite{Liu2020}, suggest that direct frictional contacts with the walls are not dominant and do not control the final arrested state of the piles. Instead, confinement primarily affects the relaxation dynamics, likely through viscous lubrication effects.

\end{document}